# Intrinsic thermal conductivity of ZrC from low to ultra-high temperatures: A critical revisit

Janak Tiwari, Tianli Feng*

Department of Mechanical Engineering, University of Utah, Salt Lake City, UT 84112, USA

**Corresponding Author:**

*tianli.feng@utah.edu

**ABSTRACT**

Current phonon transport theory based on ground-state calculations has been successful in predicting thermal conductivity at room and medium temperatures but may misrepresent behavior at high temperatures. In this work, we predict the thermal conductivity ($\kappa$) of ZrC including electronic and phonon contributions from 300 K to 3500 K, by including high-order phonon scattering, lattice expansion, temperature-dependent (TD) harmonic and anharmonic force constants, and inter-band phonon conduction by using first principles. For the phonon transport, we find that four-phonon scattering significantly reduces the phonon thermal conductivity ($\kappa_{ph}$), e.g., by ~60% and ~75% at 2500 K and 3500 K, respectively. After including four-phonon scattering and all other factors, $\kappa_{ph}$ shows a ~$T^{-1.5}$ rather than ~$T^{-1}$ dependence. The contribution from inter-band (Wigner) phonon conduction is small, even at ultra-high temperatures. The temperature dependence of anharmonic force constants decreases the phonon scattering cross-section at elevated temperatures and increases the $\kappa_{ph}$ significantly (by 52% at 3500 K). For the electronic thermal transport, we find that it is sensitive to and can be changed by 20% by the TD lattice constants. The Lorenz number varies from 1.6 to $3.3 \times 10^{-8}$ W·Ω·K$^{-2}$ at different temperatures. The theoretical prediction in the literature overpredicts $\kappa_{ph}$ (e.g., ~28%) and underpredicts the $\kappa_{el}$ (e.g., ~38%), resulting in an overall underprediction of $\kappa$ (~26% at 1500 K). The impacts of grain size and defects are found strong, and no reported experimental data has reached the intrinsic theoretical thermal conductivity of ZrC yet.

## I. INTRODUCTION

High melting point [1,2], high hardness [3], good thermal conductivity [4,5], excellent chemical stability, good electrical conductivity [5], high strength and stiffness [6,7], and resistance to oxidation even at high temperature [8] makes ZrC a suitable material for various high-temperature engineering applications [9–13] such as rocket nozzles, turbine blades, heat shields, cutting tools, refractory materials, armor materials, etc. Furthermore, ZrC is a common material used in the nuclear industry due to its ability to withstand high temperatures and radiation exposure, making it an ideal choice for fuel elements [14,15]. In addition to these applications, ZrC is also used to make electrodes in batteries and fuel cells [16,17]. To effectively design and optimize these applications, a thorough understanding of the underlying heat transfer mechanism on ZrC is necessary.

The study of heat transfer mechanisms in ZrC has been a topic of great interest. Many experimental thermal conductivity data of ZrC from room temperature up to its melting point (~3700 K) have been reported [4,5,18–23]. However, the measured thermal conductivity data are scattered across the literature. For example, at room temperature, the values vary from as low as 20 W·m$^{-1}$·K$^{-1}$ to as high as 40 W·m$^{-1}$·K$^{-1}$, and at 2000 K, they are scattered from 22 to 46 W·m$^{-1}$·K$^{-1}$. ZrC is a semi-metallic material, with the total thermal conductivity ($\kappa$) arising from the combined effect of electronic ($\kappa_{el}$) and lattice thermal conductivity ($\kappa_{ph}$). However, since direct measurement of lattice thermal conductivity is challenging, it is typically derived from the total thermal conductivity by subtracting the electronic contribution by the Wiedemann-Franz law [24,25]. This approach leaves some uncertainty because the Lorenz number ($L$) is not necessarily $L_0$= 2.44×10$^{-8}$ W·Ω·K$^{-2}$. Moreover, $\kappa_{ph}$ is found to be more sensitive to external factors such as defects, grain boundaries, and impurities than $\kappa_{el}$, making it further challenging to unveil the intrinsic thermal conductivity directly from the existing experimental data.

Different theoretical studies [26–29] based on molecular dynamics (MD) and density functional theory (DFT) have been done to understand the thermal conductivity of ZrC. Crocombette [26] studied the phonon and electronic thermal conductivity of ZrC from 1000 to 3500 K using molecular dynamics (MD). They further found that the impacts of defects are strong. Although they account for temperature effect and predict thermal conductivity correctly using MD, they did not discuss on intrinsic scattering and fundamental thermal transport mechanism. Zhou *et al.* found that carbon vacancy [27] and Hf additive [28] greatly suppresses the phonon thermal conductivity by using first principles calculations. However, therr results are based on Debye-Callaway model (of which accuracy can be questionable) and ignored the higher order scattering, which is found important at elevated temperature [30,31]. Morevoer, their study is focused only on lattice thermal conductivity. Mellan *et. al* [29] predicted the thermal conductivity of ZrC by incorporating both three phonon scatterings (3ph) and four phonon scatterings (4ph), as well as phonon renormalization for the 2nd order force constant. However, their calculations were only based on ground state force constants (GSFC), which may not be accurate for high-temperature calculations as found in Ref [32]. In this work, we predict the thermal conductivity of ZrC by including high-order phonon scattering, lattice expansion, and temperature-dependent force constants (TDFC). We also calculate the off-diagonal (Wigner) thermal conductivity, which is found to be significant for various materials at high temperatures. Additionally, we have analyzed the impact of extrinsic factors such as grain boundary scattering, defect scattering, and impurity scattering on phonon thermal conductivity at different temperatures.

## II. METHODOLOGY

ZrC is a semi-metallic material with thermal conductivity (κ) being

$$\kappa = \kappa_{ph} + \kappa_{el}. \tag{1}$$

Based on the Boltzmann transport equation (BTE), the $\kappa_{ph}$ can be calculated as:

$$\kappa_{ph}^{\alpha\beta} = \sum_{\lambda} c_\lambda v_\lambda^\alpha v_\lambda^\beta \tau_\lambda^{ph}, \tag{2}$$

where λ = (**q**,υ) denotes the phonon mode with wave vector **q** and polarization υ. $c_\lambda$ is the specific heat per mode, $v_\lambda^\alpha$ and $v_\lambda^\beta$ are phonon group velocities along α and β directions. $\tau_\lambda^{ph}$ is the phonon relaxation time and can be calculated using $\frac{1}{\tau_\lambda^{ph}} = \frac{1}{\tau_\lambda^{ph-ph}} + \frac{1}{\tau_\lambda^{ph-i}} + \frac{1}{\tau_\lambda^{ph-el}}$ where the three terms on the right are phonon-phonon (including 3ph and 4ph), phonon-isotope, and phonon-electron (ph-el) scattering rates, respectively.

Similarly, κ$_{el}$ is calculated based on BTE and Onsager [33] relations as:

$$\sigma_{\alpha\beta} = -\frac{e^2 n_s}{V} \sum_{i\mathbf{k}} \frac{df_{i\mathbf{k}}}{d\varepsilon} v_{i\mathbf{k}}^\alpha v_{i\mathbf{k}}^\beta \tau_{i\mathbf{k}}^{el} \qquad (3)$$

$$[\sigma S] = -\frac{e\, n_s}{VT} \sum_{i\mathbf{k}} (\varepsilon_{i\mathbf{k}} - \mu) \frac{df_{i\mathbf{k}}}{d\varepsilon} v_{i\mathbf{k}}^\alpha v_{i\mathbf{k}}^\beta \tau_{i\mathbf{k}}^{el} \qquad (4)$$

$$K_{\alpha\beta} = -\frac{n_s}{VT} \sum_{i\mathbf{k}} (\varepsilon_{i\mathbf{k}} - \mu)^2 \frac{df_{i\mathbf{k}}}{d\varepsilon} v_{i\mathbf{k}}^\alpha v_{i\mathbf{k}}^\beta \tau_{i\mathbf{k}}^{el} \qquad (5)$$

Here, $\sigma_{\alpha\beta}$ and $S_{\alpha\beta}$ denotes the electrical conductivity and Seeback coefficient. $K_{\alpha\beta}$ is used to calculate κ$_{el}$ as κ$_{el} = K - S\sigma ST$. $e$ is the elementary charge, **k** is the electronic wave vector at band index $i$, $V$ is the volume of the primitive cell, $f_{i\mathbf{k}}$ is Fermi-Dirac distribution. $\varepsilon_{i\mathbf{k}}$ and $\mu$ denotes electron energy and chemical potential respectively. $v_{i\mathbf{k}}$ and $\tau_{i\mathbf{k}}^{el}$ denotes electron velocity and relaxation time and $\alpha$ and $\beta$ are directional components. The formulation is referred from Refs. [33,34]

Figure 1 shows the computational workflow of the study. The first principles calculations based on DFT are performed by using Vienna Ab initio Simulation Package (VASP) [35,36], using projected augmented wave (PAW) [37] method and Perdew-Burke-Ernzerhof (PBE) [38] functional. ZrC belongs to the Fm3m space group and exhibits a cubic FCC structure. The plane-energy cutoff is 500 eV and the energy and force convergence threshold are 2×10$^{-8}$ eV and 2×10$^{-8}$ eV·Å$^{-1}$, respectively. The relaxed lattice constant is 4.710 Å, which closely resembles the experimental value [2] of 4.694 Å. Harmonic force constants (HFC) are extracted using Phonopy [39]. The 3$^{rd}$ and 4$^{th}$ order anharmonic force constants (AFC) are calculated using Thirdorder and Fourthorder packages built inside ShengBTE [40], considering the 4$^{th}$ and 2$^{nd}$ nearest atoms, respectively. In all DFT calculations, supercells of 4×4×4 (128 atoms) are used with a **k**-point grid of 6×6×6. Temperature-dependent 2$^{nd}$, 3$^{rd}$, and 4$^{th}$ order force constants are obtained by the TDEP method [41,42] with the input being the energy, forces, and stresses of atoms in randomly a displaced supercell lattice at provided temperature. 300 random configurations are found to be sufficient at lower temperatures (300 to 1000 K), while 500 configurations are needed at higher temperatures (1500 to 3500 K) to obtain converged force constants.

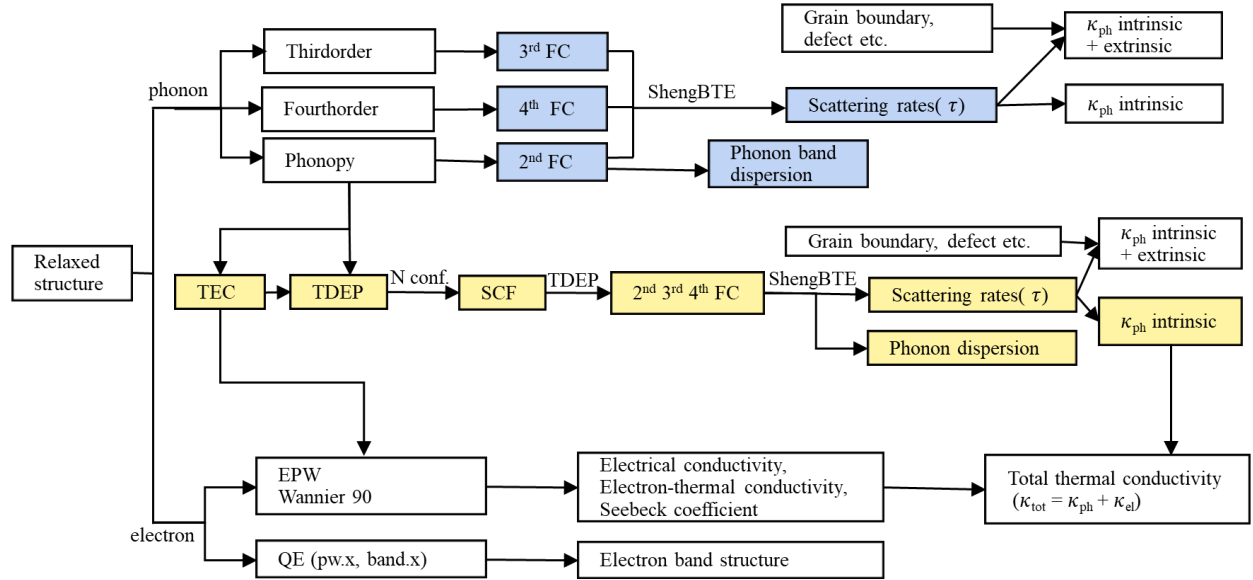

Figure 1: Computation workflow of this study.

ShengBTE [40] is used to solve the BTE and calculate the 3ph and 4ph rates and $\kappa_{ph}$. The convergence of ShengBTE is tested and provided in the Supplemental Material [43]. Specifically, 3ph and 4ph are found to converge at a **q**-mesh density of 36×36×36 and 12×12×12, respectively. Different extrinsic scattering rates, such as grain boundary scattering, defect scattering, and isotope scattering are calculated using their respective formulations, which are provided in the Supplemental Material [43]. Phonon-electron scattering is calculated using density functional perturbation theory (DFPT) [44–46] and maximally localized Wannier functions using EPW [47–50]. Finally, total scattering rates are obtained by adding all the phonon scattering rates and are used to calculate the intrinsic $\kappa_{ph}$.

Electronic contribution to $\kappa$ is calculated using EPW [47–50] interfaced with Quantum Espresso [51,52]. The structure is relaxed with a **k**-mesh of 12×12×12, a kinetic energy cutoff of 200 Ry, and Gaussian smearing with a spreading parameter of 0.002 Ry. Self-consistency is achieved using Davidson iterative diagonalization with a convergence threshold of $10^{-12}$. Electron-phonon scattering are calculated using a 12×12×12 **k**-mesh. Electron-electron scattering rates are calculated using a coarse 6×6×6 **q**-mesh and **k**-mesh and a fine 60×60×60 **q**-mesh and **k**-mesh for convergence. The sp$^3$ and d entanglement is used for C and Zr, respectively. The electrons' velocities are calculated using the Wannier90 [53] under EPW package. The Fermi window is selected so that the electron band dispersion obtained using the EPW package and DFT matches. Likewise, the outer window is selected so that it includes all the bands of interest. Finally, σ, κ$_{el}$, and S are calculated.

## III. RESULTS AND DISCUSSION

### A. Thermal Expansion Coefficient

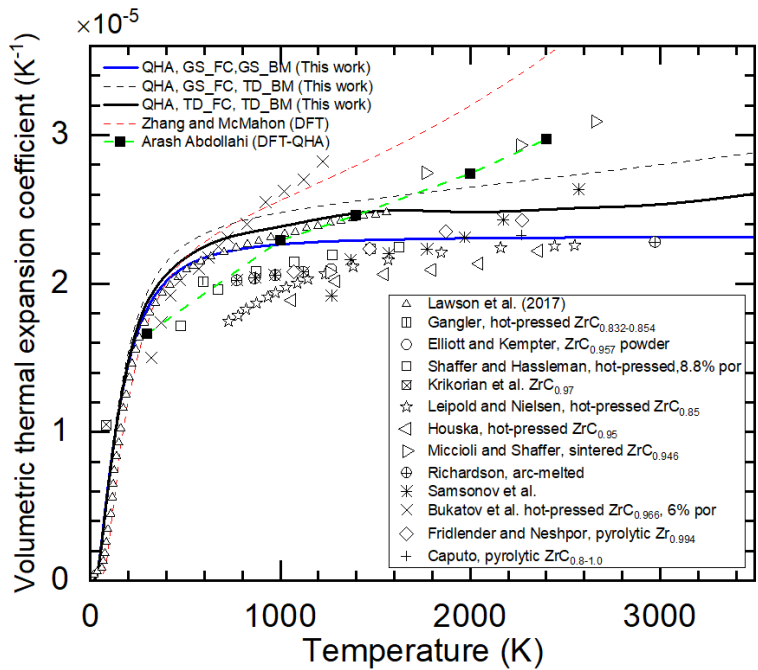

Figure 2: Predicted volumetric thermal expansion coefficient of ZrC as a function of temperature. The DFT predicted data by Zhang and McMahon [54] and Abdollahi [55] as well as different experimental data [4,23,56–58] are included for comparison.

The thermal expansion coefficient (TEC) is calculated using quasi-harmonic approximation (QHA) with the formalism found in Ref. [59]. As shown in Fig. 2, the predicted TEC (blue curve) deviates from experimental data at high temperatures, which is commonly seen for QHA. To resolve the discrepancy, we replace the constant bulk modulus (229 GPa [4]) in the formalism with temperature-dependent (TD) bulk modulus, and the predicted TEC (black dashed curve) agrees better with experimental data. After we include TDFC, the TEC (black solid curve) at high temperature agrees even better with experimental data. It is important to note that TD bulk modulus tends to increase the slope of TEC with temperature, while TDFC tends to decrease it. Overall, these findings suggest that utilizing TD bulk modulus and TDFC is a promising approach for the accurate prediction of TEC at high temperatures.

## B. Temperature-dependent phonon band dispersion

The phonon dispersion relations of ZrC at 0, 300, 1000, and 3000 K are shown in Fig. 3, which match well with the experimental data [60–62]. The temperature softening effect in ZrC is not significant. The phonon dispersion calculated from TDEP at finite temperature deviates slightly from the ground state calculations. The partial density of states (PDOS) shows that the heavier metal Zr dominates the acoustic phonon of frequency (0-10 THz), while the lighter element C dominates the optical phonon of frequency (12-20 THz). Since the acoustic phonon modes contribute the most to lattice heat transfer, we can expect Zr vibration dominates the heat transfer.

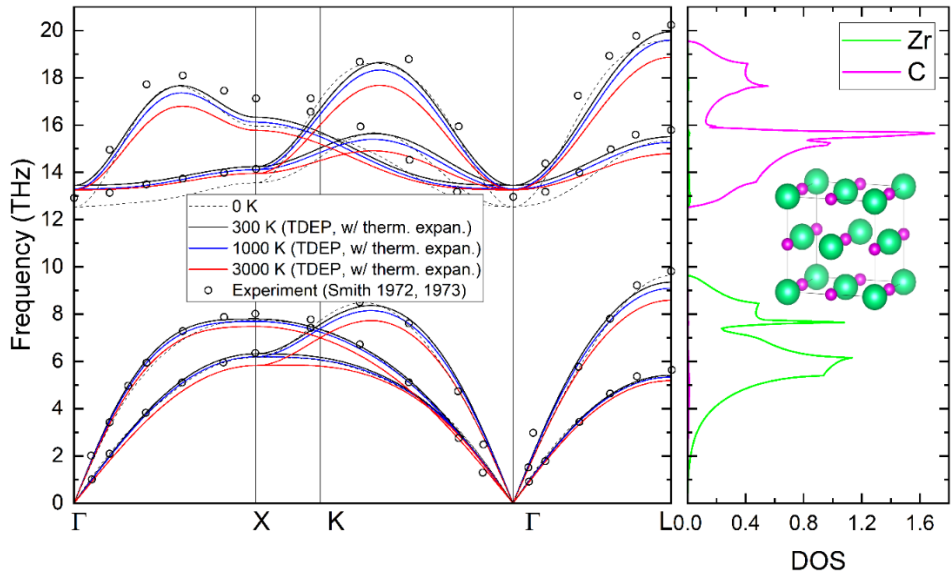

Figure 3: Phonon dispersion relations of ZrC at 0, 300, 1000, and 3000 K, compared to the experimental data [60–62]. The right side shows the partial density of states of Zr and C at 0 K.

## C. Scattering rates

Figures 4 (a), (b), and (c) show the comparison between the 3ph rates calculated using the GSFC and TDFC at 300, 1000, and 3000 K respectively. Compared to GSFC, TDFC decreases the scattering rates at all temperatures. This result is in accordance with Ref. [32], where we showed a decrease in phonon-scattering cross-section at elevated temperatures due to the softening of AFC. Similarly, Figs. (d), (e), and (f) depict the 4ph rates calculated using GSFC and TDFC at 300, 1000, and 3000 K, respectively. Interestingly, acoustic phonon scattering rates decrease while the optical phonon scattering rates increase when using the TDFC instead of the GSFC. This trend becomes more prominent as the temperature increases and is consistent with the findings for $UO_2$ [32]. The reason is still unclear. One possible reason is that optical branches are flattened more than acoustic branches, leading to larger scattering phase space.

The 3ph, 4ph, and ph-el rates, at three different temperatures of 300, 1000, and 3000 K are presented in Figs. 4 (g), (h), and (i), respectively. 3ph dominates throughout all temperatures, and 4ph becomes important only at high temperatures, while ph-el is insignificant at all temperatures (even though a very small portion of phonons show high ph-el rates). 4ph plays more important role in optical phonons than acoustic phonons, similar to the findings in Si, BAs, and diamond [30]. The large 4ph rates are due to the acoustic-optical phonon band gap, which restricts 3ph processes.

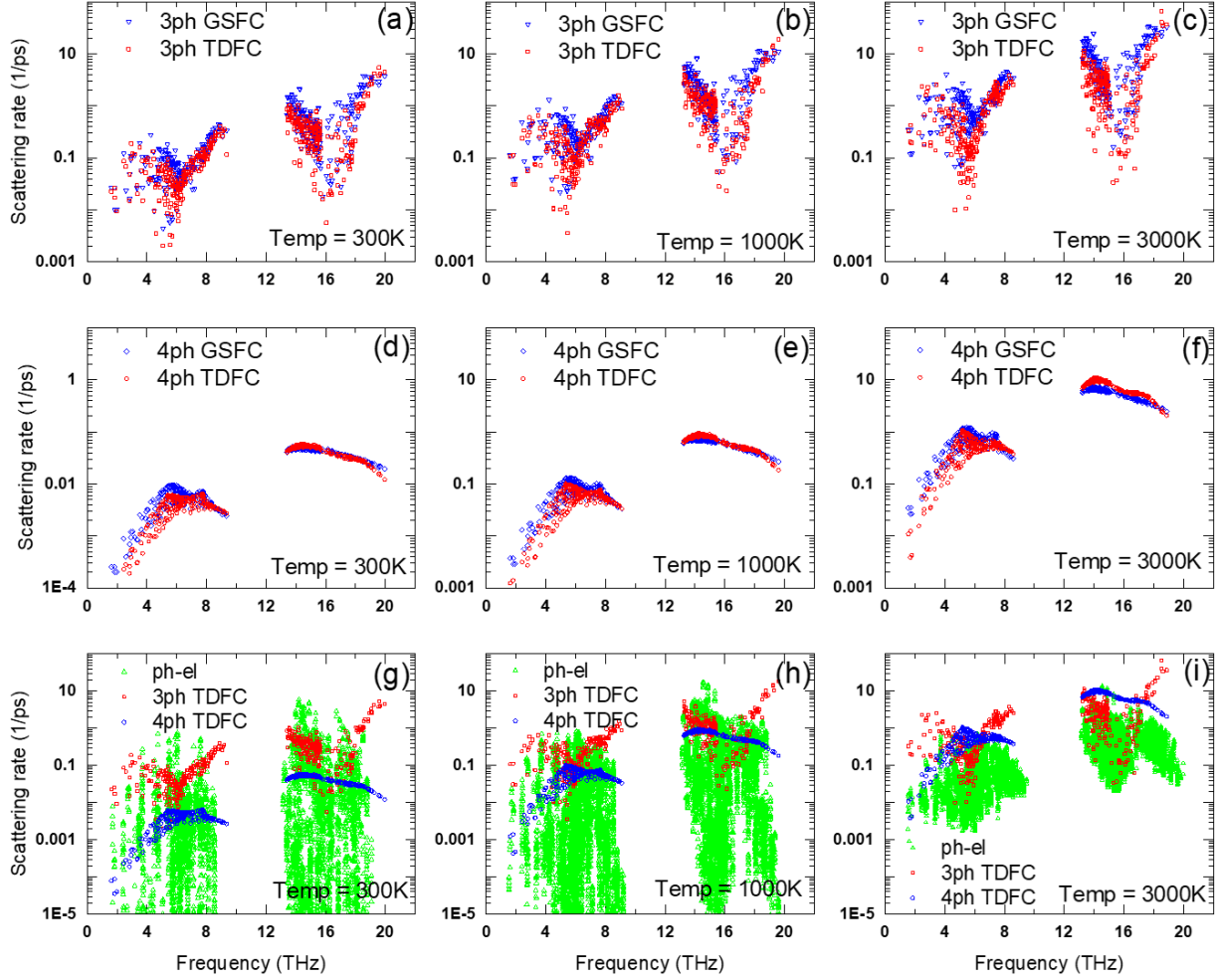

Figure 4: Comparison of 3ph (a, b, c), 4ph (d, e, f), and ph-el (g, h, i) scattering rates of ZrC at different temperatures. Results by using GSFC and TDFC are compared.

**D. Phonon thermal conductivity calculation**

Figure 5 shows the $\kappa_{ph}$ as a function of temperature, using various force constants. In the following, we track the changes of $\kappa_{ph}$ at a low (300 K) and a high temperature (3500 K) when we gradually increase the calculation comprehensivity. First, we calculate the basic 3ph thermal conductivity using GSFC, which gives 56.9 and 6.3 W·m$^{-1}$·K$^{-1}$ at 300 and 3500 K, respectively. When we replace the GS HFC with TD HFC, $\kappa_{ph}$ increases by 9% and 33% to 62.1 and 8.4 W·m$^{-1}$·K$^{-1}$, respectively. Then, we include 4ph, and $\kappa_{ph}$ decreases significantly by 2% and 75% to 61 and 2.1 W·m$^{-1}$·K$^{-1}$, respectively. After that, we replace the GS AFC with the TD AFC, and $\kappa_{ph}$ increases by 0% and 52% to 61 and 3.2 W·m$^{-1}$·K$^{-1}$, respectively. This increase in $\kappa_{ph}$ is attributed to the decrease in scattering cross-section with increasing temperature [32]. In the end, we add the ph-el scattering, and $\kappa_{ph}$ slightly decreases by 10% to 55.0 and 2.9 W·m$^{-1}$·K$^{-1}$ at 300 and 3500 K, respectively. The inset illustrates the reduction of thermal conductivity by 4ph as a function of temperature. Further analysis on 4ph using different force constants is provided in the Supplemental Material [43]. The off-diagonal term, calculated from Wigner formalism, is not found to make a significant contribution, unlike in some other materials [63,64]. Although the Wigner contribution increases with temperature, it only reaches a maximum value of 0.2 W·m$^{-1}$·K$^{-1}$ at 3500 K, which is much lower than the standard Peierl $\kappa_{ph}$ of 2.9 W·m$^{-1}$·K$^{-1}$ at that temperature. Overall, considering all these intrinsic effects, the $\kappa_{ph}$ is found to follow a temperature dependence of $\sim T^{-1.5}$ rather than $\sim T^{-1}$.

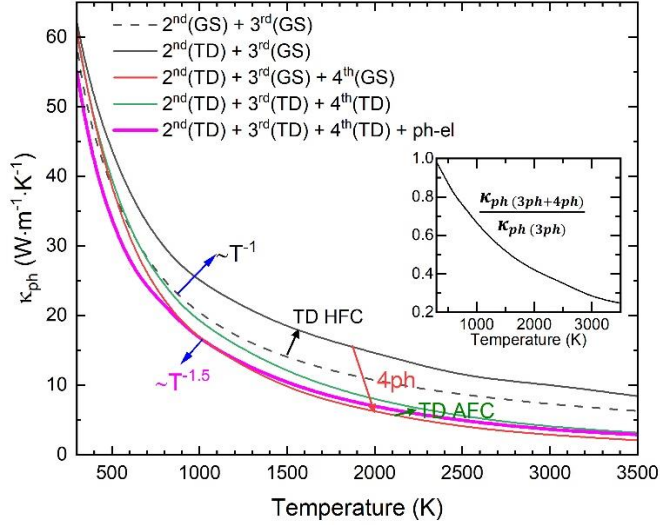

Figure 5: Temperature-dependent phonon thermal conductivity of ZrC calculated using different scattering mechanisms. The inset shows the relative contribution of 4ph rates compared to 3ph rates.

**E. Electronic thermal conductivity calculation**

The predicted electrical conductivity ($\sigma$) as a function of temperature is shown in Fig. 6. The experimental data along with the DFT prediction from Ref. [29] are also shown for comparison. The $\sigma$ decreases with temperature monotonically as a result of the increase of electron-phonon scattering [34,65–67]. The $\sigma$ calculated in this study is consistent with the results of stoichiometric

ZrC obtained in Ref. [26] using ab initio molecular dynamics simulations. However, it should be noted that our σ prediction is higher than most of the experimental data, particularly at low temperatures. This could be because the DFT calculations assume a perfect crystal and do not consider the effects of defects (especially the carbon vacancies) and porosity that present in experimental samples. This idea is supported by the findings in Ref. [26], where an increase in electrical resistivity (equivalently decrease in σ) is observed with impurities and defects in the sample. Furthermore, the Lorenz number ($L$) using the Wiedemann-Franz law: $L = \kappa_{el}/\sigma T$ is calculated, which is found to deviate significantly from the Sommerfeld value of $2.44 \times 10^{-8}$ W·Ω·K$^{-2}$ and vary from 1.6 to $3.3 \times 10^{-8}$ W·Ω·K$^{-2}$ at different temperatures. This deviation from the standard value highlights the limitations of using the Lorenz number to calculate $\kappa_{el}$ from σ, as such an approach can lead to overprediction at lower temperatures and underprediction at higher temperatures.

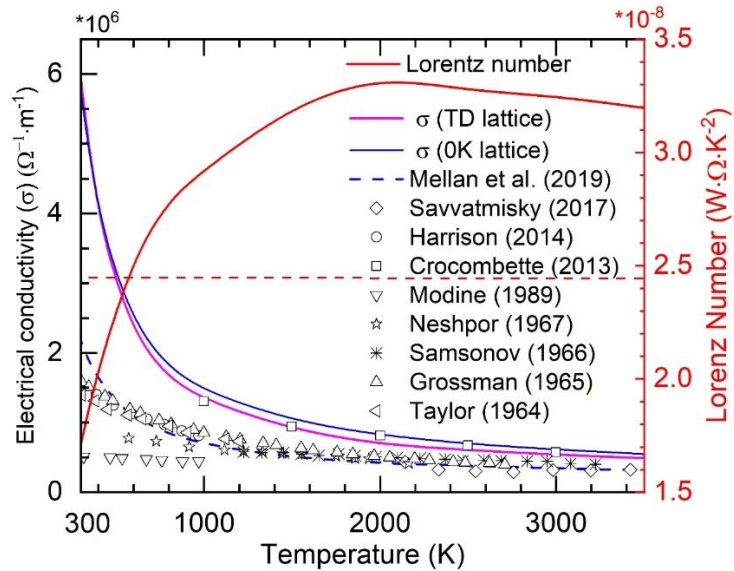

Figure 6: Electrical conductivity and Lorenz number of ZrC as a function of temperature. Literature data of electrical conductivity [2,4,5,21,22,26,29,68,69] are included for comparison.

The $\kappa_{el}$ is compared to $\kappa_{ph}$ as a function of temperature in Fig. 7. Phonons dominate thermal transport at lower temperatures whereas electrons dominate at higher temperatures, similar to results found in Refs. [26,29]. As temperature increases, e.g., from 300 K to 3500 K, $\kappa_{el}$ increases from 30 W·m$^{-1}$·K$^{-1}$ to 54.7 W·m$^{-1}$·K$^{-1}$, whereas $\kappa_{ph}$ decreases from 55.0 W·m$^{-1}$·K$^{-1}$ to 2.9 W·m$^{-1}$·K$^{-}$

[1]. At room temperature, $\kappa_{ph}$ accounts for roughly 70% of $\kappa$, which declines to just about 35% and 10% at 1000 and 3500 K respectively.

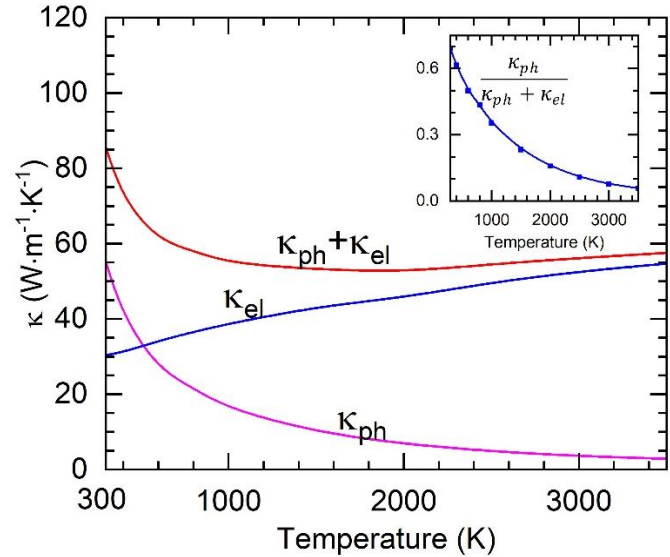

Figure 7: Temperature-dependent $\kappa_{ph}$, $\kappa_{el}$, and $\kappa$ of ZrC. The inset shows the relative contribution of $\kappa_{ph}$.

In Fig. 8, the predicted $\kappa$ (= $\kappa_{ph}$ + $\kappa_{el}$) is compared to various experimental data. The blue-shaded and yellow-shaded region represents the contribution of $\kappa_{el}$ and $\kappa_{ph}$, respectively. We can find that the experimental data are scattered and considerably lower than our prediction. This should be due to the presence of various defects, porosity, and vacancies in the experimental samples, which significantly decrease $\kappa$. Carbon vacancy is known to inevitably present in ZrC due to the intrinsic thermodynamic instability [70,71]. This decrease is much more significant at lower temperatures compared to higher temperatures due to the suppression of $\kappa_{ph}$. $\kappa_{el}$ also gets suppressed due to impurities, defects, and vacancy scattering; however, the effect is smaller as found by Crocombette [26] through molecular dynamics. In that study, the predicted $\kappa$ for stoichiometric ZrC was much higher than the reported experimental data. However, when some vacancy was introduced, both $\kappa_{el}$, and $\kappa_{ph}$ decreased, and the predicted $\kappa$ matched the experimental data. It is worth noting that our $\kappa$, as well as $\kappa_{el}$, matches Crocombette [26] results with good accuracy.

We also compare our results to those predicted in Ref. [29] by first principles. We find that they underpredict $\kappa_{el}$ and overpredict $\kappa_{ph}$. As a result, their total $\kappa$ prediction agrees with ours at low and ultra-high temperatures, but is lower than ours in the intermediate temperature range, from 500 to 3000 K. In Ref. [29], at the temperature of 1500 K, $\kappa_{ph}$ is overestimated by ~27.5% (9.93→~12.67 W·m$^{-1}$·K$^{-1}$), and $\kappa_{el}$ is underestimated by ~38% (43.21 → ~26.78 W·m$^{-1}$·K$^{-1}$), leading to κ underestimation of ~25.81% (53.15 → ~39.43 W·m$^{-1}$·K$^{-1}$). At ultra-high temperatures of 3500 K, $\kappa_{ph}$ is overestimated by ~75% and $\kappa_{el}$ is underestimated by ~8%, leading to a slight κ underestimation of ~4%. At lower temperatures of 300 K, $\kappa_{ph}$ is overestimated by ~26.5% and $\kappa_{el}$ is underestimated by ~50.6%, leading to a slight κ underestimation of ~1.0%. The matching of the predicted κ in Ref. [29] with experimental data is due to the error-cancellation effect where the prediction underpredicts the κ and the experimental samples' defects reduce κ.

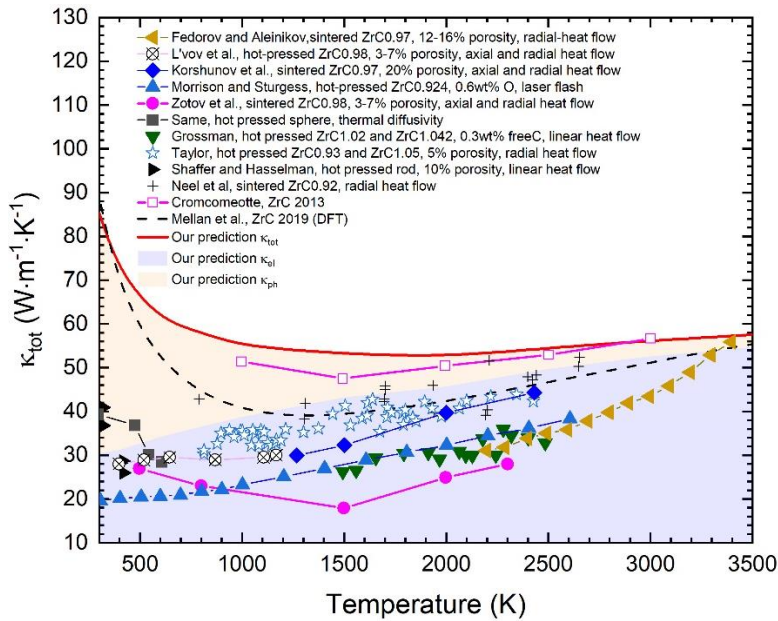

Figure 8: Comparison of our predicted thermal conductivity with various literature data collected by Jackson and Lee [4], and theoretical prediction by Crocombette [26] (ab initio molecular dynamics), and Mellan et. al. [29] (DFT).

**F. Effect of extrinsic defects and grain boundary**

To understand the effect of extrinsic factors such as grain size, we calculate the cumulative thermal conductivity with respect to the mean free path (MFP). As shown in Fig. 9 (a), in bulk ZrC, 80% of $\kappa_{ph}$ at 300, 1000, and 3000 K is contributed by the phonons with an MFP of less than 300, 60, and 20 nm, respectively. The effect of grain boundaries on thermal conductivity is more pronounced at lower temperatures. The MFP of electrons is much smaller than that of phonons. 80% of heat transfer occurs with electrons having MFPs less than 15, 4, and 0.7 nm. This suggests that the contribution of grain boundary scattering is much less significant for electrons compared to phonons. Figure 9 (b) shows the normalized $\kappa_{ph}$ as a function of grain size from 10 to 10000 nm. The solid line represents the current method (TDFC, 3ph+4ph), while the dashed line represents the traditional approach (GSFC, 3ph). The graph explains how much $\kappa_{ph}$ gets suppressed (in terms of %) at various grain sizes and temperatures. For instance, if the grain boundary size is 100 nm, the $\kappa_{ph}$ at 300, 1000, 2000, and 3000 K is roughly 55%, 76%, 87%, and 94% of the bulk values, respectively. The inset shows the grain size required to reduce $\kappa_{ph}$ to be 80% of the bulk values. At 300 and 3500 K, such grain sizes are 445 and 25 nm, respectively. The difference in grain boundary scattering between TDFC and GSFC is not substantial. Since the grain size in the experimental samples [2,4] is in the order of µm ( ~2 to 20 µm), the grain boundary scattering might not be the primary reason behind $\kappa_{ph}$ suppression.

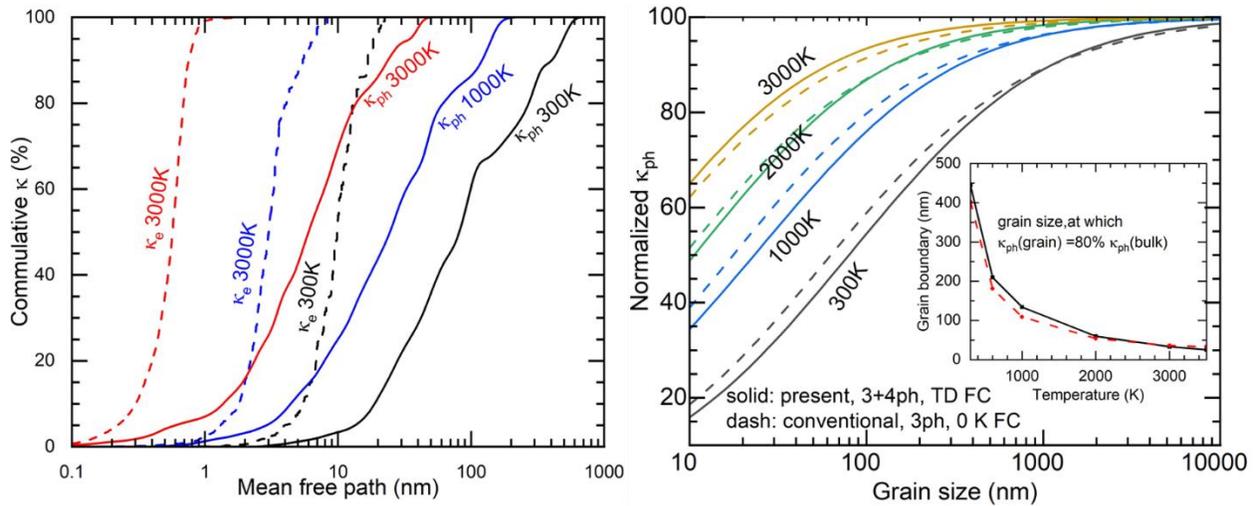

Figure 9: (a) Cumulative $\kappa_{ph}$ and $\kappa_{el}$ with mean free path at various temperatures. (b) Percentage normalized $\kappa_{ph}$ as a function of grain size. The inset shows the grain size at which $\kappa_{ph}$ is 80% of the bulk value.

The impact of carbon and zirconium vacancies on $\kappa_{ph}$ of ZrC is calculated and found to be strong, using perturbation theory, as shown in Fig. 10 (a). For example, at 300 K, 1% and 2% carbon vacancies can decrease $\kappa_{ph}$ by 39% and 56% respectively. The impact of Zr vacancy is even stronger, e.g., at 300 K, even 0.5% of Zr vacancy can lead to a decrease in $\kappa_{ph}$ by 80%. This can be explained by the partial density of states presented in Fig. 3. Zr has a much larger mass than C and dominates the acoustic frequency bands, which dominate the heat transfer. A small concentration of Zr vacancy can lead to a significant decrease in $\kappa_{ph}$.

Figure 10 (b) displays κ of ZrC for various C/Zr ratios, with experimental data collected by Jackson and Lee [4] and DFT prediction by Mellan et al. [29]. The figure reveals that the experimental κ increases exponentially as the crystal structure approaches the stoichiometric ratio, which is in line with the trend predicted by DFT. It is worth noting that our prediction accounts only for the $\kappa_{ph}$ suppression caused by C and Zr vacancies. If we include the $\kappa_{ph}$ and $\kappa_{el}$ suppression due to impurities, porosity, and grain boundaries as well, we can safely assume that our prediction will become lower and approaches the experimental value. This suggests that the experimental κ value does not reach the maximum theoretical limit of κ of ZrC, as $\kappa_{ph}$ and $\kappa_{el}$ are suppressed by the presence of vacancies, impurities, grain boundaries, and porosity in the sample.

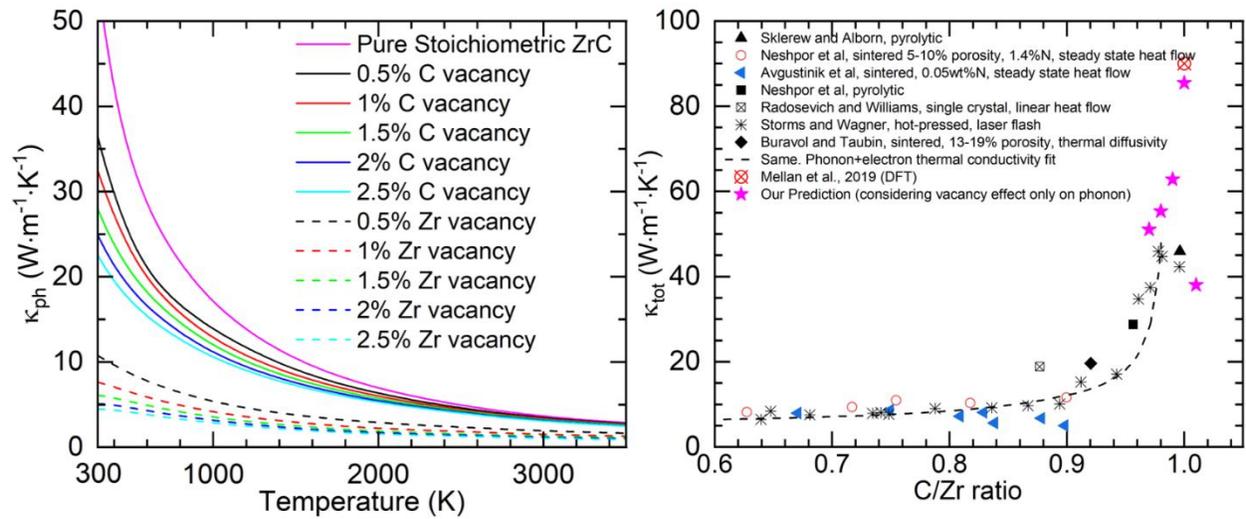

Figure 10: (a) Effect of C and Zr vacancy on phonon thermal conductivity of ZrC. (b)Variation of total thermal conductivity of ZrC with different C/Zr ratios. DFT prediction by Mellan *et al.* [29] and various experimental data collected by Jackson and Lee [4] are presented for comparison.

## IV. CONCLUSIONS

In this study, the thermal transport of ZrC is predicted using the first principles calculations. Various factors affecting thermal transport, especially at high temperatures, are considered, including high-order phonon scattering, lattice expansion, TDFC, and inter-band phonon conduction. The following conclusions can be drawn. (1) TD bulk modulus and TDFC are important for the accurate prediction of TEC at high temperatures. (2) Phonons and electrons dominate the heat transfer at lower and higher temperatures respectively. $\kappa_{ph}$ contributes roughly 70% at room temperature and declines to about 35% and 10% when the temperature increases to 1000 and 3500 K, respectively. (3) The 4ph is important at high temperatures and reduces $\kappa_{ph}$ by 2%, 34%, 59%, and 76% at 300, 1000, 2000, and 3500 K, respectively. (4) The impact of electron-phonon scattering increases with temperature but is much smaller than 3ph and 4ph scattering. (5) Temperature-dependent AFC decreases the phonon scattering cross-section at elevated temperatures and increases the $\kappa_{ph}$ significantly (by 52% at 3500 K). (6) The maximum Wigner off-diagonal (diffuson) contribution to $\kappa_{ph}$ (0.2 W·m$^{-1}$·K$^{-1}$) is much lower compared to standard Peierls phonon contribution $\kappa_{ph}$ (2.9 W·m$^{-1}$·K$^{-1}$ at 3500 K). (7) The $\sigma$ decreases monotonically while $\kappa_{el}$ increases as temperature increases. The Lorenz number ($L$) deviates significantly from the Sommerfeld value of $L_0 = 2.44 \times 10^{-8}$ W·$\Omega$·K$^{-2}$, highlighting the limitations of using the $L_0$ to calculate $\kappa_{el}$. (8) No experimental data have reached the predicted intrinsic thermal conductivity values due to the presence of inherent defects in the experimental samples. The matching of $\kappa$ predicted in the literature with experimental data is due to the error-cancellation effect where the prediction underpredicts the $\kappa$ and the experimental samples' defects reduce $\kappa$. Overall, our study makes a critical revisit to the thermal transport from room to ultra-high temperatures.

**ACKNOWLEDGMENTS**

This work is supported by the National Science Foundation (NSF) (award number: CBET 2212830). The computation used resources from the Center for High Performance Computing (CHPC) at the University of Utah, the Advanced Cyberinfrastructure Coordination Ecosystem: Services & Support (ACCESS) of NSF, and the National Energy Research Scientific Computing Center (NERSC), a DOE Office of Science User Facility under Contract No. DE-AC02-05CH11231 and award No. BES-ERCAP0022132.

**Data availability**

Source data are provided along with this paper. All other data that support the plots within this paper are available from the corresponding authors on reasonable request.

**Code availability**

The codes used in this study are available from the corresponding authors upon request.

**Author contributions**

T.F. conceived the idea and guided the project. J.T performed the simulations and wrote the original manuscript. J.T and T.F. both revised the manuscript.

**Competing interests**

The authors declare no competing interests.

**Additional information**

Correspondence and requests for materials should be addressed to T.F.